\documentclass[prd,preprintnumbers,nofootinbib,aps,9pt]{revtex4}
\usepackage{subeqnarray}
\usepackage{epsfig,amsmath,amssymb}
\usepackage{mathrsfs}
\usepackage[usenames,dvipsnames]{color}
\usepackage[pagebackref=true, colorlinks=true]{hyperref}
\definecolor{redish}{rgb}{0.7,0.2,0.0}  % color defined in (r=red,g=green,b=blue) model
\definecolor{bluish}{rgb}{0.2,0.5,0.8}
\hypersetup{linkcolor=redish,          % color of internal links
                  citecolor=blue,        % color of links to bibliography
                  filecolor=magenta,      % color of file links
                  urlcolor=bluish}          % color of the url
\DeclareFontFamily{U}{rsfs}{}         % Formal Script            %
\DeclareFontShape{U}{rsfs}{m}{n}{<5> rsfs5 <6><7> rsfs7          %
  <8><9><10><10.95><12><14.4><17.28><20.74><24.88> rsfs10}{}     %
\DeclareMathAlphabet{\mathfs}{U}{rsfs}{m}{n}                     %
\newcommand{\mfs}[1]{\mathfs {#1}}                              %

\newcommand{\ba}{\nopagebreak[3]\begin{eqnarray}}
\newcommand{\ea}{\end{eqnarray}}
\newcommand{\bii}{\begin{itemize}}
\newcommand{\eii}{\end{itemize}}
\newcommand{\nn}{\nonumber}

\newcommand{\sO}{{\mfs O}}

\newcommand{\f}{\frac}

\def \b{\beta}
\def \l{\ell}
\def \g{\gamma}
\def \e{\epsilon}
\def \lp{\l_p}
\def \j{\sqrt{j(j+1)}}

\def \lm{\lambda}

\def \G{\mathcal{G}}

\def \sj{s_j^{\star}}

\def \th{\theta}
\def \tn{\theta_0}

\def \O{\Omega}

\def \({\left(}
\def \){\right)}
\def \[{\left[}
\def \]{\right]}
\begin{document}
\title{Conformal blocks on a 2-sphere with indistinguishable punctures and implications on black hole entropy}
\author{Abhishek Majhi}%
\email{abhishek.majhi@gmail.com}
\affiliation{Instituto de Ciencias Nucleares\\
Universidad Nacional Autonoma de Mexico\\
A. Postal 70-543, Mexico D.F. 04510, Mexico}

\pacs{}
\begin{abstract}
The dimensionality of the Hilbert space of a Chern-Simons theory on a 3-fold, in the presence of Wilson lines carrying spin representations, had been counted by using its link with the Wess-Zumino theory, with level $k$, on the 2-sphere with points (to be called punctures) marked by the piercing of the corresponding Wilson lines and carrying the respective spin representations. It is shown, in the weak coupling (large $k$) limit, the formula decouples into two characteristically distinct parts; one mimics the dimensionality of the Hilbert space of a collection of non-interacting spin systems and the other is an effective overall correction contributed by all the punctures. The exact formula yield from this counting has been shown earlier to have resulted from the  consideration of the punctures to be distinguishable.  We investigate the same counting problem by considering the punctures to be indistinguishable. Although the full formula remains undiscovered, nonetheless, we are able to impose the relevant statistics for indistinguishable punctures in the approximate formula resulting from the weak coupling limit. As an implication of this counting, in the context of its relation  to that of black hole entropy calculation in quantum geometric approach, we are able to show that the logarithmic area correction, with a coefficient of $-3/2$,  that results in this method of entropy calculation, in independent of whether the punctures are distinguishable or not.
\end{abstract}
\maketitle

\section{Introduction}
The dimension of the Hilbert space of the Chern-Simons(CS) theory on a 3-fold with boundary 2-sphere is given by the number of conformal blocks of an $SU(2)_k$ Wess-Zumino(WZ) theory of level $k$ on that boundary\cite{wit}. In particular, in the presence of Wilson lines carrying spin representations in the 3-fold, which pierce the boundary 2-sphere producing marked points (henceforth, to be called punctures) carrying the corresponding spin representations, the dimension of the CS theory is then given by the number of conformal blocks of the WZ theory on the 2-sphere with punctures. This counting was explicitly done in \cite{km98} to yield this number $\Omega(j_1,j_2,\cdots,j_p)$ (say) for a set of $p$ punctures carrying spins  $j_1,j_2,\cdots,j_p$.

Now, if one wishes to count this number for all possible spins, then one does a sum over all possible spins i.e. $\sum_{j_1,\cdots,j_p}\Omega(j_1,j_2,\cdots,j_p)$, which can be recast as a sum over all possible spin configurations  i.e. $\sum_{\{s_j\}}\Omega[\{s_j\}]$ by the use of multinomial expansion (a spin configuration $\{s_j\}$ constitutes a set such that there are $s_j$ number of spin $j$)\cite{ampm2}. This yields a formula for $\Omega[\{s_j\}]$ which manifests the number of conformal blocks of the WZ theory on a punctured sphere with spin configuration $\{s_j\}$, if the punctures are considered to be distinguishable. 

Here, we  investigate this counting problem by considering the punctures to be indistinguishable. Unfortunately, in this case, we do not have a way, unlike the multinomial expansion, to go over from a set of spins to spin configurations simply because we do not know how to take the sum over spins\footnote{This particular problem even exists in the elementary statistical mechanics of a collection of particles\cite{pathria}.}. Also, the way in which the counting exercise originates in terms of the fusion matrices of the WZ theory \cite{km98} it is not clear at the moment how to differentiate between the counting methods for distinguishable and indistinguishable punctures. So, we do an approximation  to the formula for $\Omega(j_1,\cdots,j_p)$  and express it in a convenient form where the imposition of the statistics, depending on the disnguishability of the punctures, becomes manifestly trivial. Consequently, we arrive at an effective formula for the $\Omega[\{s_j\}]$ for indistinguishable punctures. An immediate application of this approximate result is found in the context of black hole entropy calculation in the quantum geometric approach. We show that the subleading logarithmic area correction of the entropy is independent of whether the punctures are distinguishable or not.  %However, the exact formula for $\Omega[\{s_j\}]$ for indistinguishable punctures derived from the scratch using the fusion matrices of WZ theory, still remains unknown.

\section{The weak coupling limit}
The number of conformal blocks of an $SU(2)_k$ WZ theory on a 2-sphere with a set of $p$ number of punctures, carrying spins $j_1,j_2,\cdots,j_p$, is given by\cite{km98}
\ba
\Omega(j_1,j_2,\cdots,j_p)%&=&\f{2}{k+2}\sum_{r=0}^{k/2}\f{\prod_{l=1}^p\sin\[\f{(2j_l+1)(2r+1)\pi}{k+2}\]}{\sin\[\f{(2r+1)\pi}{k+2}\]^{p-2}}\nn\\
&=&\f{2}{k+2}\sum_{a=1}^{k+1}\f{\prod_{l=1}^p\sin\[\f{(2j_l+1)a\pi}{k+2}\]}{\sin\[\f{a\pi}{k+2}\]^{p-2}}\label{main}
\ea
where $0\leq j_l\leq k/2$, for all $l=1,\cdots,p$ and one has to use the Verlinde formula (see e.g. \cite{cftbook}) to obtain this expression. Now, we can approximate the formula (\ref{main}) in the limit $k\to \infty$ and $p\to\infty$ as follows. Defining a variable $\th=a\pi/(k+2)$, in the limit $k\to\infty$,  eq.(\ref{main}) can be recast in the following integral form
\ba
%g[\{s_j\}]&\simeq&\f{2}{k+2}\int^{k+1}_{1}\sin^2\f{a\pi}{k+2}\prod_{j}\left\{\f{\sin\f{a\pi(2j+1)}{k+2}}{\sin\f{a\pi}{k+2}}\right\}^{s_j} da\nn\\
\Omega(j_1,j_2,\cdots,j_p)&=&\f{2}{\pi}\int^{\pi -\e}_{\e}\sin^2\th ~\prod_{l=1}^p\left\{\f{\sin (2j_l+1)\th}{\sin \th}\right\}~ d\th\label{eq2}
\ea
where $\e=\pi/(k+2)$. Further, for $k\to\infty$, $\e\to 0$ and hence, eq.(\ref{eq2}) can be rewritten with approximated limits of integration  as
\ba
\Omega(j_1,j_2,\cdots,j_p)&=&\f{2}{\pi}\int^{\pi }_{0}\sin^2\th ~\prod_{l=1}^p\left\{\f{\sin (2j_l+1)\th}{\sin \th}\right\}~ d\th\nn\\
&=&\f{1}{\pi}\int^{\pi}_{0}\exp\left[G(\th) \right]d\th - \f{1}{\pi}\int^{\pi}_{0}\exp\left[\ln(\cos 2\th)+G(\th) \right]d\th\label{sadd7}
\ea
where $G(\th)=\sum_{l=1}^p\log\left\{\f{\sin (2j_l+1)\th}{\sin \th}\right\}$. The above two integrations can be performed by the saddle point method, similar to the one performed in ref.\cite{ampm2}. The saddle point comes out to be at $\th=\th_0=0$ for  both $G(\th)$ and $\ln(\cos 2\th)+G(\th)$. Then, Taylor expanding both these two terms around $\th_0$ up to the second order i.e.$(\th-\th_0)^2$ term and neglecting the higher order terms, the integrals can be performed to arrive at the following expression:
\ba
\Omega(j_1,j_2,\cdots,j_p)\simeq\f{1}{\sqrt{2\pi}}\prod_{l=1}^p(2j_l+1)\left\{\sqrt{\f{1}{\alpha}}~\text{Erf}\left[\pi\sqrt{\alpha/2}\right]-\sqrt{\f{1}{4+\alpha}}~\text{Erf}\left[\pi\sqrt{(4+\alpha)/2}\right]\right\}\label{re}
\ea
where $\alpha=\f{4}{3}\sum_{l=1}^p ~ j_l(j_l+1)$ and ``Erf[.]'' stands for {\it error function}\cite{math}. Now, for large $p~(\gg 1)$, which we shall take as the limit $p\to\infty$, we have  $\alpha\to\infty$. Further, plotting the function $\text{Erf}\left[\pi\sqrt{\alpha/2}\right]$ with $\alpha$ one can see that the function attains the constant value of unity for $\alpha\to\infty$ and consequently, one can also find that $\lim_{\alpha\to\infty}\text{Erf}\left[\pi\sqrt{(4+\alpha)/2}\right]=\lim_{\alpha\to\infty}\text{Erf}\left[\pi\sqrt{\alpha/2}\right]= 1$. Consequently, the $\alpha$-dependent term within the braces in eq.(\ref{re}) simplifies to  $\left\{\alpha^{-1/2}-(4+\alpha)^{-1/2}\right\}$. As $\alpha\to\infty$, the leading order contribution in the asymptotic expansion of this term comes out to be $\alpha^{-3/2}$. Hence,  eq.(\ref{re}) reduces to the following:
\ba
\Omega(j_1,j_2,\cdots,j_p)%&\simeq&\f{C}{\sqrt{2\pi}}\prod_{l=1}^p(2j_l+1)\left(\sqrt{\f{1}{\alpha}}~-\sqrt{\f{1}{4+\alpha}}\right)\nn\\
&\sim&\f{1}{\sqrt{2\pi}}~\alpha^{-3/2}\prod_{l=1}^p(2j_l+1)\label{scset}.
\ea

At this point it is necessary to point out the fact that the $\alpha$-dependent part in eq.(\ref{re}), and hence in eq.(\ref{scset}), results from the second order term in the saddle point approximation of eq.(\ref{sadd7}), whereas the $\alpha$-independent part results from the zeroth order term. The implication of this statement will be clear while we explore an application of this result in the context of black hole entropy later.

It may be noted that the above calculations, although is similar to the one carried out in \cite{ampm2}, is actually slightly different in the details because the present one is done with a set of spins $(j_1,\cdots,j_p)$ whereas the one in \cite{ampm2} was done with a spin configuration $\{s_j\}$. However, this being only the difference at the mathematical level, there lies a deeper conceptual motivation behind doing this approximation with a spin sequence. One may note that the way in which the calculations have been done in ref.\cite{ampm2}, {\it implicitly assumes a priori that the punctures are distinguishable}. It leaves no way to address the counting problem for indistinguishable punctures which is the issue of interest in the present context. But, having done the above calculation with a set of spins $(j_1,\cdots,j_p)$, we will now be able to address the issue of indistinguishable punctures. This is because the formula (\ref{main}) takes a very particular form given by eq.(\ref{scset}), which is suitable for implementing the statistics associated with a collection of non-interacting spin systems, as it happens in elementary statistical mechanics. We explain this point in detail as follows.

Neglecting the irrelevant constant $1/\sqrt{2\pi}$, the point to be noted about the formula (\ref{scset}) is that, it has manifestly decoupled into two characteristically distinct parts:
\begin{itemize}
\item  $\prod_{l=1}^p(2j_l+1)$: This piece mimics the dimensionality of the Hilbert space of an ordinary collection of $p$ number of non-interacting spin carrying entities ( i.e. $\bigotimes_{l=1}^p{\cal H}_l$, where ${\cal H}_l$ is the Hilbert space associated with an individual spin carrying entity). Henceforth, we shall address it as $dim[\bigotimes_{l=1}^p{\cal H}_l]$. Needless to say, it is  trivial to see that the statistical effect on the counting due to nature of the punctures will be encoded in this part of the formula. 
\item $\alpha^{-3/2}$: This piece is an overall effective correction produced by the collective impact of all the spins. To be precise, the quantity $\alpha=\f{4}{3}\sum_l j_l(j_l+1)$ is nothing but the sum of a function of the spin values of the punctures. This is independent of whether the punctures are distinguishable or not. 
\end{itemize}
Further,  one can check from \cite{ampm2} that if we carry out the whole program of approximation from the beginning in terms of spin configuration, considering distinguishable punctures, then $\alpha$ becomes proportional to $\sum_js_j j(j+1)$. It is again trivial to see in this form that $\alpha$ is the sum of a function of the spin values of the punctures and  can not get affected by the nature of punctures.
Due to the above fact, whenever we want to estimate $\Omega[\{s_j\}]$, only the piece $dim[\bigotimes_{l=1}^p{\cal H}_l]$, will be affected by the nature of the punctures. Consequently, it becomes quite easy to foresee what $\Omega[\{s_j\}]$ will be in case of distinguishable and indistinguishable punctures.  We elaborate that as follows.

{\bf Distinguishable punctures:} Taking into account all these above facts about the formula (\ref{scset}) and considering the punctures to be distinguishable, we have 
\ba
\Omega[\{s_j\}]=\f{1}{\sqrt{2\pi}} \alpha^{-3/2}\f{p!}{\prod_js_j!}\prod_j(2j+1)^{s_j}\label{disapp}
\ea
where $p:=\sum_js_j$ is the total number of punctures. 
One can arrive at the above equation just by  summing over all spins
in the eq.(\ref{scset}) followed by the use of multinomial expansion to arrive at an equation manifesting sum over spin configurations. Alternatively, one can see the eq.(\ref{scset}) and ask the question that if we have a spin configuration $\{s_j\}$ instead of a set $(j_1,\cdots,j_p)$, then what the formula will look like. Then one imposes Maxwell-Boltzmann counting to arrive at the above result (which is possible due to the decoupling that we explained earlier). 

{\bf Indistinguishable punctures:} However, if we consider that the punctures are indistinguishable, then we do not have a recipe such as `sum over all spins' in this case. The only way is to ask the question again that if we have a spin configuration $\{s_j\}$ instead of a set $(j_1,\cdots,j_p)$, then what the formula will look like\footnote{ Indeed this is the same thing which is done in elementary statistical mechanics involving a collection of non-interacting spin systems e.g. ideal gas \cite{pathria}.}. The answer is the following:
\ba
\Omega[\{s_j\}]=\f{1}{\sqrt{2\pi}} \alpha^{-3/2}\prod_j\f{(s_j+2j)!}{s_j!(2j)!}\label{indisapp}
\ea
This is simply obtained by implementing the Bose-Einstein(BE) method of counting. The reason behind the implementation of the BE statistics (why not Fermi-Dirac) is that, there can be 
any number of punctures with any spin value since there is no physical restriction (unlike Pauli exclusion principle for Fermions).

%Eq.(\ref{indisapp}) is a generic result for the space of conformal blocks  of an $SU(2)_k$ WZ theory on a 2-sphere with a large number of indistinguishable punctures having a spin configuration $\{s_j\}$ in the large $k$ limit. Hence, it is applicable 

\section{Logarithmic correction to black hole entropy from indistinguishable punctures}
%It is explicit from the calculations shown above that the saddle point approximation gives rise to the factor of $\alpha^{-3/2}$ from the $\xi^2$ term in the exponent of the integrand in eq.(\ref{sadd}) and the zeroth order term is the part $dim[\bigotimes_{l=1}^p{\cal H}_l]$. The formula given in eq.(\ref{indisapp}), which is obtained from eq.(\ref{sadd}) by invoking Bose-Einstein statistics for indistinguishable punctures and consequently inherits the same characteristic feature, has an interesting implication in the black hole entropy calculation.

The link between 3-dimensional CS theory and 2-dimensional WZ theory makes the whole counting problem relevant for state counting for black hole horizons in quantum geometry\cite{km98} and this makes eq.(\ref{indisapp}) to yield an interesting result in this particular context. In the quantum geometric approach to black hole entropy calculation it has been recently shown how the area law (i.e. $S=A/4\lp^2$, $\lp$ being the Planck length and $A$ being the horizon area) can emerge from the zeroth order term for the indistinguishable punctures i.e. eq.(\ref{indisapp}) without the $\alpha$-correction\cite{amindis}. The entropy has been calculated in \cite{amindis} by using the method of most probable distribution and hence, is just given by $S\simeq \log\Omega[\{\sj\}]$, where $\sj$ is the most probable distribution of spins. So, it is quite easy to see that, if we consider the extra factor of $\alpha^{-3/2}$, the  entropy will now be given by
\ba
S=\f{A}{4\lp^2}-\f{3}{2}\log \alpha^* + \cdots\label{edicorr}
\ea 
where 
\ba
\alpha^*=\f{4}{3}\sum_j\sj j(j+1)\label{alstar}.
\ea
 Since $\alpha$ results from a second order approximation, the $\sj$ that should be used to calculate $\alpha^*$ must result from a calculation involving the zeroth order term of $\Omega[\{s_j\}]$ in eq.(\ref{indisapp}) (i.e. neglecting the $\alpha$-correction). This $\sj$ has already been calculated in \cite{amindis} and is given by 
\ba
\sj= \f{2j}{e^{\b\j}-1}\label{sjstar}
\ea
where $\b$ is a parameter which takes a very small value for the area law to follow\footnote{Strictly speaking $\b=8\pi\lm\g$ where $\lm$ and $\g$ are two parameters, both of which have been shown to take small values for the area law to follow \cite{amindis}. However, in the present case it will suffice to consider only the parameter $\b$ and consider that it is very small.}. 

To debrief, eq.(\ref{sjstar}) results from the variation of $\Omega[\{s_j\}]$ given in eq.(\ref{indisapp}), but without the $\alpha$-correction, with respect to $s_j$, subject to the constraint 
\ba
C:~~ 8\pi\g\lp^2\sum_js_j\j=A.
\ea
To mention, $8\pi\g\lp^2\sum_js_j\j$ is the quantum geometric area of the sphere with punctures having spin distribution $s_j$ and $\g$ is free parameter of the theory which is determined by demanding that the entropy be given by the area law in the leading order.  The constraint $C$ defines the microcanonical area ensemble. For the details of these calculations one can look into \cite{amindis}. 

Now, it is straightforward to estimate $\alpha^*$ as follows. Since $\b$ takes a very small value, the leading order term in the expansion of the right hand side of eq.(\ref{sjstar}) in terms of $\b$ is given by
\ba
\sj\simeq\f{2j}{\b\j}\label{sjstarapp}.
\ea
%To see this one can just check that $\lim_{x\to0}(e^x-1)^{-1}=x^{-1}+$~smaller terms.
Hence, using eq.(\ref{sjstarapp}) in eq.(\ref{alstar}) we have 
\ba
\alpha^*&=&\f{4}{3}\sum_j\f{2j}{\b\j}j(j+1)\nn\\
&=& \f{1}{3\pi\b\g}\(\f{A}{\lp^2}\)\label{alstarf}
\ea
where one has to use also the fact that $\sj$ satisfies the constraint $C$. Hence, using eq.(\ref{alstarf}) in eq.(\ref{edicorr}), we have 
\ba
S=\f{A}{4\lp^2}-\f{3}{2}\log \(\f{A}{\lp^2}\) + \cdots\label{logind}
\ea
It may be mentioned that $\b$ and $\g$ asymptotes to constant values for large area $(A/\lp^2\gg\sO(1))$, which is the limit -- i) in which all these calculations are performed ~ ii) consistent with the limit $k\to\infty$ as it is defined as $k=A/4\pi\g\lp^2$ in the quantum geometric framework\cite{qg2}. Thus, the effects of logarithmic terms involving those parameters are irrelevant in the entropy calculation. To conclude, from eq.(\ref{logind}) and refs.\cite{ampm2,sigma}, we are able to show that the sub-leading term in the black hole entropy, in quantum geometry, is independent of whether the punctures on the horizon are distinguishable or not.

\section{Conclusion}
 In a nutshell, we have estimated the dimensionality of the space of conformal blocks of an $SU(2)_k$ Wess-Zumino theory, of level $k$, on a 2-sphere, with indistinguishable punctures, which is a boundary of a 3-fold having a Chern-Simons theory on it, in presence of Wilson lines. However, we have estimated the result only in an approximate sense in the large $k$ limit. %Although the exact estimate still goes begging, we hope that this approximate formula proves to be useful in some associated physics problems. 
 Further we discuss an immediate implication of our results in the context of black hole entropy calculation in the quantum geometric approach. Since the counting problem discussed here is directly related to the microstate count of black holes in quantum geometry,  we are able to show that the logarithmic correction viz. $-3/2\log(A/\lp^2)$, which results from the entropy calculation in this quantum geometric approach, is independent of whether the spin-carrying punctures on the 2-sphere cross-section of the black hole horizon are distinguishable or not. However, we have imposed the statistics for the indistinguishable punctures only in an approximate sense and the full structure of the formula remains hitherto unknown. It will be interesting to find some way to discover the full formula, as we think that the estimate about the space of conformal blocks of $SU(2)_k$ Wess-Zumino theory on a 2-sphere with indistinguishable spin-carrying punctures may find it applications, apart from the black hole entropy calculation which has been discussed here, in other fields of physics related to the application of two dimensional conformal field theory.

\vspace{0.1cm}
~\\
{\bf Acknowledgments :} This work is funded by DGAPA-UNAM project IG100316.

\end{document}